\documentclass[aps,prl,preprint,superscriptaddress]{revtex4-1}

\usepackage{graphicx}
\usepackage{color}
\usepackage{longtable}

\begin{document}


  \title{Search for bosonic superweakly interacting massive dark matter particles with the XMASS-I detector}



  \author{K.~Abe}
  \affiliation{Kamioka Observatory, Institute for Cosmic Ray Research,
    the University of Tokyo, Higashi-Mozumi, Kamioka, Hida, Gifu, 506-1205, Japan}
  \affiliation{Kavli Institute for the Physics and Mathematics
    of the Universe (WPI), the University of Tokyo, Kashiwa,
    Chiba, 277-8582, Japan}

  \author{K.~Hieda}
  \affiliation{Kamioka Observatory, Institute for Cosmic Ray Research,
    the University of Tokyo, Higashi-Mozumi, Kamioka, Hida, Gifu, 506-1205, Japan}

  \author{K.~Hiraide}
  \affiliation{Kamioka Observatory, Institute for Cosmic Ray Research,
    the University of Tokyo, Higashi-Mozumi, Kamioka, Hida, Gifu, 506-1205, Japan}
  \affiliation{Kavli Institute for the Physics and Mathematics
    of the Universe (WPI), the University of Tokyo, Kashiwa,
    Chiba, 277-8582, Japan}

  \author{S.~Hirano}
  \affiliation{Kamioka Observatory, Institute for Cosmic Ray Research,
    the University of Tokyo, Higashi-Mozumi, Kamioka, Hida, Gifu, 506-1205, Japan}

  \author{Y.~Kishimoto}
  \affiliation{Kamioka Observatory, Institute for Cosmic Ray Research,
    the University of Tokyo, Higashi-Mozumi, Kamioka, Hida, Gifu, 506-1205, Japan}
  \affiliation{Kavli Institute for the Physics and Mathematics
    of the Universe (WPI), the University of Tokyo, Kashiwa,
    Chiba, 277-8582, Japan}

  \author{K.~Ichimura}
  \affiliation{Kamioka Observatory, Institute for Cosmic Ray Research,
    the University of Tokyo, Higashi-Mozumi, Kamioka, Hida, Gifu, 506-1205, Japan}
  \affiliation{Kavli Institute for the Physics and Mathematics
    of the Universe (WPI), the University of Tokyo, Kashiwa,
    Chiba, 277-8582, Japan}

  \author{K.~Kobayashi}
  \affiliation{Kamioka Observatory, Institute for Cosmic Ray Research,
    the University of Tokyo, Higashi-Mozumi, Kamioka, Hida, Gifu, 506-1205, Japan}
  \affiliation{Kavli Institute for the Physics and Mathematics
    of the Universe (WPI), the University of Tokyo, Kashiwa,
    Chiba, 277-8582, Japan}

  \author{S.~Moriyama}
  \affiliation{Kamioka Observatory, Institute for Cosmic Ray Research,
    the University of Tokyo, Higashi-Mozumi, Kamioka, Hida, Gifu, 506-1205, Japan}
  \affiliation{Kavli Institute for the Physics and Mathematics
    of the Universe (WPI), the University of Tokyo, Kashiwa,
    Chiba, 277-8582, Japan}

  \author{K.~Nakagawa}
  \affiliation{Kamioka Observatory, Institute for Cosmic Ray Research,
    the University of Tokyo, Higashi-Mozumi, Kamioka, Hida, Gifu, 506-1205, Japan}

  \author{M.~Nakahata}
  \affiliation{Kamioka Observatory, Institute for Cosmic Ray Research,
    the University of Tokyo, Higashi-Mozumi, Kamioka, Hida, Gifu, 506-1205, Japan}
  \affiliation{Kavli Institute for the Physics and Mathematics
    of the Universe (WPI), the University of Tokyo, Kashiwa,
    Chiba, 277-8582, Japan}

  \author{H.~Ogawa}
  \affiliation{Kamioka Observatory, Institute for Cosmic Ray Research,
    the University of Tokyo, Higashi-Mozumi, Kamioka, Hida, Gifu, 506-1205, Japan}
  \affiliation{Kavli Institute for the Physics and Mathematics
    of the Universe (WPI), the University of Tokyo, Kashiwa,
    Chiba, 277-8582, Japan}

  \author{N.~Oka}
  \affiliation{Kamioka Observatory, Institute for Cosmic Ray Research,
    the University of Tokyo, Higashi-Mozumi, Kamioka, Hida, Gifu, 506-1205, Japan}

  \author{H.~Sekiya}
  \affiliation{Kamioka Observatory, Institute for Cosmic Ray Research,
    the University of Tokyo, Higashi-Mozumi, Kamioka, Hida, Gifu, 506-1205, Japan}
  \affiliation{Kavli Institute for the Physics and Mathematics
    of the Universe (WPI), the University of Tokyo, Kashiwa,
    Chiba, 277-8582, Japan}

  \author{A.~Shinozaki}
  \affiliation{Kamioka Observatory, Institute for Cosmic Ray Research,
    the University of Tokyo, Higashi-Mozumi, Kamioka, Hida, Gifu, 506-1205, Japan}

  \author{Y.~Suzuki}
  \affiliation{Kamioka Observatory, Institute for Cosmic Ray Research,
    the University of Tokyo, Higashi-Mozumi, Kamioka, Hida, Gifu, 506-1205, Japan}
  \affiliation{Kavli Institute for the Physics and Mathematics
    of the Universe (WPI), the University of Tokyo, Kashiwa,
    Chiba, 277-8582, Japan}

  \author{A.~Takeda}
  \affiliation{Kamioka Observatory, Institute for Cosmic Ray Research,
    the University of Tokyo, Higashi-Mozumi, Kamioka, Hida, Gifu, 506-1205, Japan}
  \affiliation{Kavli Institute for the Physics and Mathematics
    of the Universe (WPI), the University of Tokyo, Kashiwa,
    Chiba, 277-8582, Japan}

  \author{O.~Takachio}
  \affiliation{Kamioka Observatory, Institute for Cosmic Ray Research,
    the University of Tokyo, Higashi-Mozumi, Kamioka, Hida, Gifu, 506-1205, Japan}

  \author{D.~Umemoto}
  \affiliation{Kamioka Observatory, Institute for Cosmic Ray Research,
    the University of Tokyo, Higashi-Mozumi, Kamioka, Hida, Gifu, 506-1205, Japan}

  \author{M.~Yamashita}
  \affiliation{Kamioka Observatory, Institute for Cosmic Ray Research,
    the University of Tokyo, Higashi-Mozumi, Kamioka, Hida, Gifu, 506-1205, Japan}
  \affiliation{Kavli Institute for the Physics and Mathematics
    of the Universe (WPI), the University of Tokyo, Kashiwa,
    Chiba, 277-8582, Japan}

  \author{B.~S.~Yang}
  \affiliation{Kamioka Observatory, Institute for Cosmic Ray Research,
    the University of Tokyo, Higashi-Mozumi, Kamioka, Hida, Gifu, 506-1205, Japan}
  \affiliation{Kavli Institute for the Physics and Mathematics
    of the Universe (WPI), the University of Tokyo, Kashiwa,
    Chiba, 277-8582, Japan}

  \author{S.~Tasaka}
  \affiliation{Information and Multimedia Center, Gifu University, Gifu 501-1193, Japan}

  \author{J.~Liu}
  \affiliation{Kavli Institute for the Physics and Mathematics
    of the Universe (WPI), the University of Tokyo, Kashiwa,
    Chiba, 277-8582, Japan}

  \author{K.~Martens}
  \affiliation{Kavli Institute for the Physics and Mathematics
    of the Universe (WPI), the University of Tokyo, Kashiwa,
    Chiba, 277-8582, Japan}

  \author{K.~Hosokawa}
  \affiliation{Department of Physics, Kobe University, Kobe, Hyogo 657-8501, Japan}

  \author{K.~Miuchi}
  \affiliation{Department of Physics, Kobe University, Kobe, Hyogo 657-8501, Japan}

  \author{A.~Murata}
  \affiliation{Department of Physics, Kobe University, Kobe, Hyogo 657-8501, Japan}

  \author{Y.~Onishi}
  \affiliation{Department of Physics, Kobe University, Kobe, Hyogo 657-8501, Japan}

  \author{Y.~Otsuka}
  \affiliation{Department of Physics, Kobe University, Kobe, Hyogo 657-8501, Japan}

  \author{Y.~Takeuchi}
  \affiliation{Department of Physics, Kobe University, Kobe, Hyogo 657-8501, Japan}
  \affiliation{Kavli Institute for the Physics and Mathematics
    of the Universe (WPI), the University of Tokyo, Kashiwa,
    Chiba, 277-8582, Japan}

  \author{Y.~H.~Kim}
  \author{K.~B.~Lee}
  \author{M.~K.~Lee}
  \author{J.~S.~Lee}
  \affiliation{Korea Research Institute of Standards and Science, Daejeon 305-340, South Korea}

  \author{Y.~Fukuda}
  \affiliation{Department of Physics, Miyagi University of Education, Sendai, Miyagi 980-0845, Japan}

  \author{Y.~Itow}
  \affiliation{Solar Terrestrial Environment Laboratory, Nagoya University, 
    Nagoya, Aichi 464-8602, Japan}
  \affiliation{Kobayashi-Masukawa Institute for the Origin of Particles and the Universe, Nagoya University, Furu-cho, Chikusa-ku, Nagoya, Aichi, 464-8602, Japan}

  \author{K.~Masuda}
  \author{H.~Takiya}
  \author{H.~Uchida}
  \affiliation{Solar Terrestrial Environment Laboratory, Nagoya University, 
    Nagoya, Aichi 464-8602, Japan}

  \author{N.~Y.~Kim}
  \author{Y.~D.~Kim}
  \affiliation{Department of Physics, Sejong University, Seoul 143-747, South Korea}

  \author{F.~Kusaba}
  \author{K.~Nishijima}
  \affiliation{Department of Physics, Tokai University, Hiratsuka,
    Kanagawa 259-1292, Japan}

  \author{K.~Fujii}
  \author{I.~Murayama}
  \author{S.~Nakamura}
  \affiliation{Department of Physics, Faculty of Engineering, Yokohama National University, Yokohama, Kanagawa 240-8501, Japan}

  \collaboration{XMASS collaboration}
  \noaffiliation

  \date{\today}

  \begin{abstract}
    Bosonic superweakly interacting massive particles (super-WIMPs)
    are a candidate for warm dark matter.
    With the absorption of such a boson by a xenon atom
    these dark matter candidates would 
    deposit an energy equivalent to their rest mass in the detector. 
    This is the first direct detection experiment
    exploring the vector super-WIMPs in the mass range
    between 40 and 120\,keV.
    Using 165.9\,days of data no significant excess
    above background
    was observed in the fiducial mass of 41\,kg.
    The present limit for the vector super-WIMPs
    excludes the possibility that such particles 
    constitute all of dark matter.
    The absence of a signal also provides the most stringent direct
    constraint on the coupling constant of pseudoscalar
    super-WIMPs to electrons.
    The unprecedented sensitivity was achieved exploiting
    the low background at a level
    $10^{-4}$\,kg$^{-1}$keV$_{ee}^{-1}$day$^{-1}$ in the detector.
  \end{abstract}


  \maketitle


  There is overwhelming evidence for the existence of dark matter
  in the Universe. 
  Since all the evidence is gravitational, the nature of
  dark matter is not well constrained and various models have been 
  considered.
  For a model to be falsifiable in a direct detection experiment it needs to 
  allow 
  at least one interaction beyond the gravitational one. 
  A well motivated model that guides most experimental searches imagines the 
  dark matter particle as a weakly interacting thermal relic, candidates for 
  which are provided by various extensions of the standard model of particle 
  physics. 
  In the case that dark matter is such a weakly interacting massive 
  particle (WIMP), thermal decoupling after the big bang
  automatically ensures the right relic 
  abundance to account for the observed dark matter.
  Such a WIMP fits the cold dark matter (CDM) paradigm. 

  On the other hand, simulations based on this CDM scenario expect a richer 
  structure on galactic scales than
  those observed.
  Furthermore, there is so far no evidence
  of supersymmetric particles at LHC
  and therefore it is important to investigate various
  types of dark matter candidates. These facts strengthen an interest
  to consider lighter and more weakly interacting 
  particles 
  like
  super-WIMPs,
    a warm dark matter candidate \cite{Pospelov2008, Redondo2009}.
  If the mass of the super-WIMPs is above $\sim$3\,keV,
  there is no conflict with
  structure formation in the Universe \cite{WDM}.
  Bosonic super-WIMPs are
  experimentally
    interesting
  since their absorption in a
    target material
  would deposit an energy 
  essentially equivalent to the super-WIMP's rest mass.

    Here we present direct detection limits obtained
    with the XMASS-I liquid xenon detector
    for the vector and the pseudoscalar case.
    For the vector super-WIMPs search,
    this is the first direct detection experiment.
    The mass range of this study is restricted to 40-120\,keV.
    At the low mass end we are limited by increasing background,
    and at high masses the calculation in Ref.\ \cite{Pospelov2008}
    are limited to the mass of the boson less than 100\,keV.

  The absorption of a vector boson
  is very similar to the photoelectric effect when the photon
  energy $\omega$ is replaced by the vector boson mass $m_V$
  and the coupling constant is scaled appropriately.
  The cross section therefore becomes
    \cite{Pospelov2008}
  \begin{eqnarray}
    \frac{\sigma_{\rm abs}v}{\sigma_{\rm photo}(\omega=m_V)c} \approx
    \frac{\alpha'}{\alpha},
  \end{eqnarray}
  where $\sigma_{\rm abs}$ is the absorption cross section
  of the vector bosons on an atom, $v$ is the velocity
  of the incoming vector boson, $\sigma_{\rm photo}$ is the
  cross section for the photoelectric effect,
  $\alpha$ is the fine structure
  constant, and $\alpha'$ is the vector boson analogue
  to the fine structure constant. 
  For a single atomic species of atomic mass $A$,
  the counting rate
    $S_v$
  in the detector becomes
    \cite{Pospelov2008}
  \begin{eqnarray}
    S_v \approx \frac{4\times 10^{23}}{A} \frac{\alpha'}{\alpha}
    \left(\frac{\rm keV}{m_V}\right)
    \left(\frac{\sigma_{\rm photo}}{\rm barn}\right) \rm kg^{-1}day^{-1},
  \end{eqnarray}
  where the standard local dark matter density of 0.3\,GeV/cm$^3$ 
  \cite{PDG} is used.
  Valid ranges for the couplings and masses of thermally produced 
  super-WIMPs are calculated in Refs.\ \cite{Pospelov2008,Redondo2009}.

  The cross section of the axioelectric effect for the pseudoscalar 
  on the other hand is
  \begin{eqnarray}
    \frac{\sigma_{\rm abs}v}{\sigma_{\rm photo}(\omega=m_a)c} \approx
    \frac{3m_a^2}{4\pi\alpha f_a^2},
  \end{eqnarray}
  where $m_a$ is the mass of the pseudoscalar particle,
  and $f_a$ is a dimensionful coupling constant.
  The counting rate
    $S_a$
  now becomes
    \cite{Pospelov2008}
  \begin{eqnarray}
    S_a \approx \frac{1.2\times 10^{19}}{A} g_{aee}^{2}
    \left(\frac{m_a}{\rm keV}\right)
    \left(\frac{\sigma_{\rm photo}}{\rm barn}\right) \rm kg^{-1}day^{-1},
  \end{eqnarray}
  where $g_{aee} = 2m_e/f_a$, with $m_e$ being the electron mass.

XMASS-I is a large single phase
liquid-xenon detector \cite{XMASS2} located underground
(2700\,m water equivalent) at the Kamioka Observatory in Japan. 
An active target of 835\,kg of liquid xenon is held inside of a 
pentakis-dodecahedral copper structure that holds 642 inward-looking 
photomultiplier tubes (PMTs) on its approximately spherical inner surface. 
The detector is calibrated
regularly by
inserting $^{57}$Co and $^{241}$Am sources along the central 
vertical axis of the detector. 
Measuring with the $^{57}$Co source from the center of the detector volume 
the photoelectron yield
is determined to be 13.9\,photoelectrons (p.e.)/keV$_{ee}$ \cite{FN1},
where the subscript $ee$ refers to the customary electron equivalent
energy deposit.
This large photoelectron yield is realized
since the photocathode area covers $>$62\% of the inner
wall with large quantum efficiency of $\sim$30\% \cite{XMASS2}.
Data acquisition is triggered if ten or more
PMTs have signals larger than 0.2\,p.e.\ within 200\,ns.
Each PMT signal is digitized with
charge and timing resolution of
0.05\,p.e.\ and 0.4\,ns, respectively \cite{SKNIM}.
The liquid-xenon detector is located at the center of a water Cherenkov
veto counter, which is 11\,m high and has 10\,m diameter.
The veto counter is equipped with 72 50\,cm PMTs.
Data acquisition for the veto counter is triggered if eight or more of its 
PMTs register a signal within 200\,ns.
XMASS-I is the first direct detection dark matter experiment
equipped with 
such
an active water Cherenkov shield.

For both, vector and pseudoscalar type super-WIMPs,
Monte Carlo (MC) signals are generated by 
injecting
gamma rays uniformly over the 
entire active volume with a gamma energy corresponding
to the rest mass of 
the boson \cite{XMASS3}.
This procedure exploits the experimentally
relevant aspect that all the energy of a boson including
its mass given to an electron is identical
to that for gamma rays at these low energies,
albeit with different coupling constants in Eqs.\ (1) and (3).

In the present analysis we scale the observed number of 
photoelectrons by 1/13.9 to obtain an event energy
$E$
in keV$_{ee}$,
without applying the nonlinearity correction of
scintillation light efficiency.
The MC
simulation
includes this
nonlinearity of the
scintillation response \cite{XMASS2} as well as
corrections derived from the detector calibrations.
The absolute energy scale of the MC is adjusted at 122\,keV.
The systematic difference of the energy scale
between data and MC due to imperfect modeling of the nonlinearity in MC
is estimated as 3.5\% by comparing $^{241}$Am data to MC.
The decay constants of scintillation light
and the timing response of PMTs are
modeled to reproduce the
time
distribution observed with the 
$^{57}$Co (122\,keV) and $^{241}$Am (60\,keV) gamma ray sources \cite{XMASS4}.
The group velocity of the scintillation light in liquid xenon is
calculated from the refractive index ($\sim$11\,cm/ns for 175\,nm) 
\cite{LXeSpeed}.

Data taken 
in the commission phase
between December 24, 2010 and May 10, 2012
were used for the present analysis.
We selected 
the periods of operation
under what we designate normal 
data taking
conditions with a stable 
temperature
(174$\pm 1.2$\,K)
and pressure (0.160-0.164\,MPa absolute).
We have further removed the periods of operation
with excessive PMT noise,
unstable pedestal levels, or abnormal trigger rates.
Total livetime is 165.9 days.

Event selection proceeds in four stages that
we refer to as cut-1 through cut-4.
Cut-1 requires that no outer detector
trigger is associated with the events,
that they are separated from the nearest event
in time by at least 10\,ms, and that the RMS spread
of the inner detector hit timings contributing
to the trigger is less than 100\,ns. These criteria
eliminate events that are electronics or detector
artifacts rather than physical interactions
in the detector. Their application reduces
the total effective lifetime to 132.0 days in the final sample.

As discussed in Ref.\ \cite{XMASS3, XMASS1},
the main source of background
to the physics analyses stems from surface background,
especially the radioactive contaminants
in the aluminum seal of the PMTs.
We used three additional cuts to reduce those backgrounds.
Cut-2 makes use of an event vertex reconstruction.
This reconstruction is based on a maximum
likelihood evaluation of the observed light distribution
in the detector.
More detail can be found in Ref.\ \cite{XMASS2}.
We select events from the fiducial
volume by requiring that the radial distance $R$
of their reconstructed vertex from the center of
the detector is smaller than the fiducial volume radius.

The remaining two cuts deal with the issue of
mis-reconstructed events. In particular radioactive decays
on the inner surfaces of the detector pose a problem
since light emitted from the flat areas between the
PMTs is not necessarily detected by those PMTs surrounding
the emission point. Two cuts were developed to
identify and eliminate such events.
Cut-3 uses the time difference $\delta T_m$
between the first hit in an event and the mean
of the timings of the second half of all
the time-ordered hits in the event.
Events with smaller $\delta T_m$ are less likely
to be mis-reconstructed surface events and are kept.
Cut-4 eliminates events that reflect
their origin within groves or crevices in the inner
detector surface through a particular illumination pattern:
The rims of the grove or crevice restrict direct light
into a disk that is projected as a ``band'' of higher photon
counts onto the inner detector surface.
This band is characterized by the ratio
$f=$(p.e.\ in a band of width 15\,cm)/(Total p.e.\ in the event)
and $F_B$ is defined by the maximum of $f$ \cite{XMASS4}.
Events with smaller $F_B$ are less likely to originate from crevices
and are selected.
\begin{figure}
\begin{center}
\includegraphics[width=11cm]{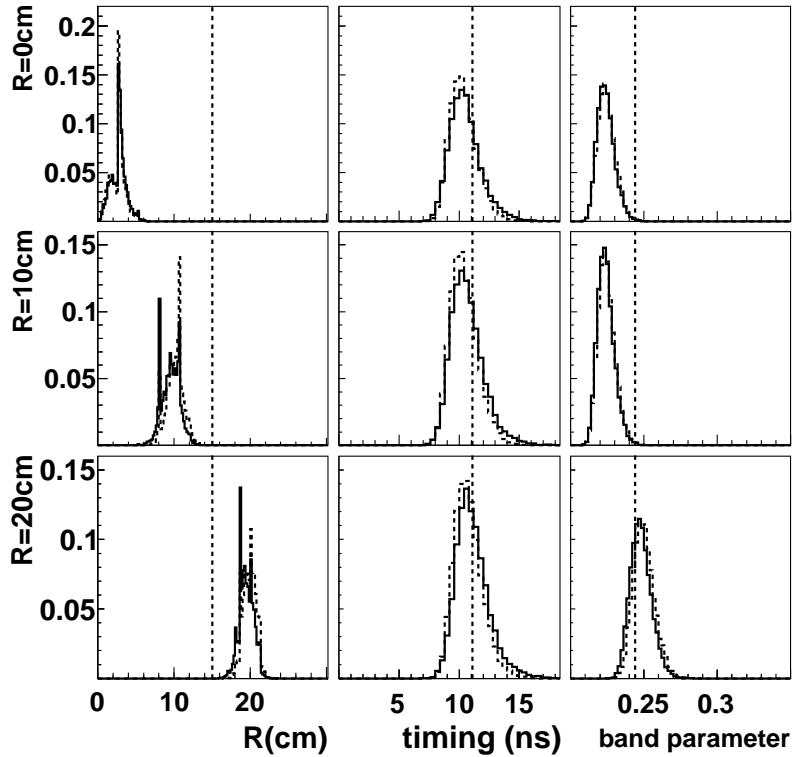}
\end{center}
\caption{Comparisons of 122\,keV gamma ray data (solid histograms)
with simulation (dashed histograms) at three positions
in the detector, $R=0$\,cm, 10\,cm, and 20\,cm, from the top
to bottom rows. From left to right, distributions for
$R$, $\delta T_m$, and $F_B$ are shown (see the text).
Data and simulation are in reasonable agreement.
The vertical dashed lines show the cut values for 120\,keV.
}
\label{xmass:datavsmc}
\end{figure}
Figure \ref{xmass:datavsmc} shows the distributions of the cut
variables
described above for $^{57}$Co source data 
and the respective simulations.
Similar distributions for $^{241}$Am
can be seen in Ref.\ \cite{XMASS4}.
The reasonable agreement demonstrates the validity of the simulation.

To maximize the sensitivity,
cut values are
optimized for each super-WIMP mass using its respective 
super-WIMP MC simulation. The optimization was done
by maximizing the ratio of the number of expected signal events
to the number of observed background events just outside
the signal range.
The signal window is $\pm 15$\,keV$_{ee}$
around the nominal masses
$m_b=m_V$ or $m_a$
shifted according to the energy scale
based on MC where the nonlinearity of the scintillation 
yield is taken into account.
Independent of the mass value this signal window contains at least
99\% of the signal. For details see Tab.\ \ref{xmass:tab}.
The number of observed background event is counted
in the energy range inside $m_b\pm 60$\,keV$_{ee}$
but outside $m_b\pm 20$\,keV$_{ee}$.
To avoid too small an acceptance,
the range
of cut values of cut-2
was restricted
in the optimization process to be larger than
15\,cm.
Table \ref{xmass:tab} summarizes the resulting cut values.
In this table, the efficiency for each cut
to retain signals
is also shown.
These efficiencies were calculated by taking the ratio between
the number of generated signal events inside a 15\,cm sphere
and the number of remaining events after the reduction.
The MC events were produced through the entire
active volume of the detector.

\begin{table*}
\begin{center}
\caption{Optimized cuts for several cases of $m_b$.
Columns $R$, $\delta T_m$, and $F_B$ list the chosen cut values.
For events with $R$, $\delta T_m$, and $F_B$ those smaller
than corresponding cut values are kept.
Column $E$ shows the range of the signal window in keV$_{ee}$.
Signal efficiencies are obtained from the detector simulation,
by taking the ratio between the number of events
in the hatched histogram in Fig.\ \protect{\ref{xmass:data}}
and the number of events generated in the fiducial mass, 41\,kg,
inside the radius 15\,cm of the detector.
The number of observed events within the signal window
is listed in the `obs.' column.
The last two columns show the resulting constraints on
$\alpha'/\alpha$ and $g_{aee}$ at 90\% CL.
}
\label{xmass:tab}
\begin{tabular}{cccccccccc}
\hline
$m_b$ (keV) & $R$ (cm) & $\delta T_m$ (ns) & $F_B$ & $E$ (keV$_{ee}$) & eff.\ (\%) & obs. & $^{214}$Pb expected & $\alpha'/\alpha$ & $g_{aee}$ \\
\hline
40 & $<$15 & $<$12.62 & $<$0.258 & 23.7--53.7& 51$\pm 13$ & 48 & $7.9\pm 0.7$ & $8.0\times 10^{-26}$ & $1.3\times 10^{-12}$\\
60 & $<$15 & $<$12.54 & $<$0.248 & 46.9--76.9 & 63$\pm 16$ & 12 & $11.6\pm 1.0$ & $6.8\times 10^{-26}$ & $8.0\times 10^{-13}$\\
80 & $<$15 & $<$11.51 & $<$0.246 & 68.1--98.1 & 59$\pm 18$ & 8 & $9.6\pm 0.8$ & $1.6\times 10^{-25}$ & $9.2\times 10^{-13}$\\
100 & $<$15 & $<$11.14 & $<$0.244 & 89--119 & 65$\pm 20$ & 15 & $11.4\pm 1.0$ & $6.0\times 10^{-25}$ & $1.4\times 10^{-12}$\\
120 & $<$15 & $<$11.11 & $<$0.244 & 111--141 & 74$\pm 23$ & 18 & $14.4\pm 1.1$ & $1.2\times 10^{-24}$ & $1.7\times 10^{-12}$\\
\hline
\end{tabular}
\end{center}
\end{table*}

\begin{figure}
\begin{center}
\includegraphics[width=11cm]{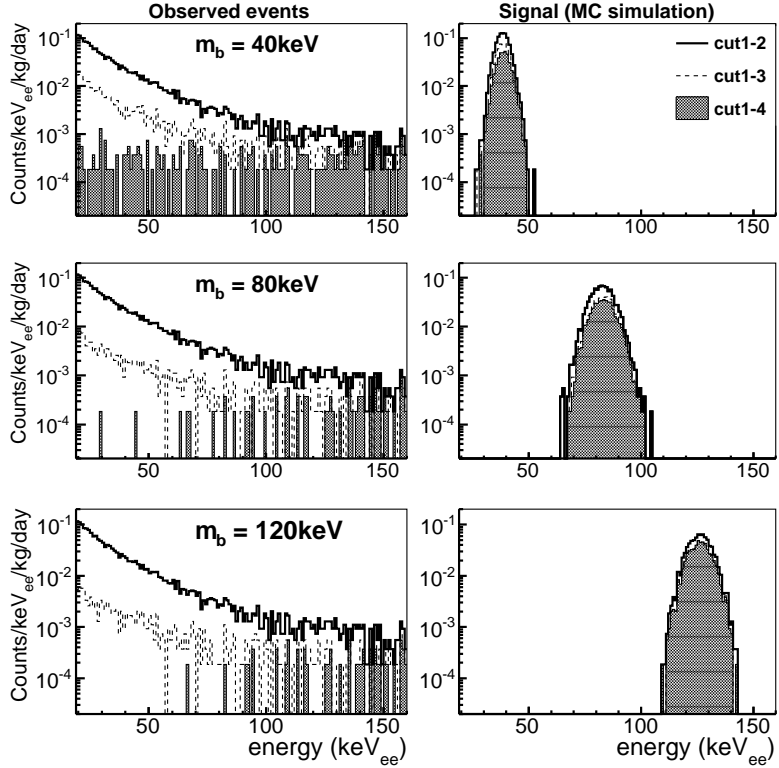}
\end{center}
\caption{Energy distribution of the observed events
(left column)
and simulated events
(right column)
remaining after each step of the cuts
optimized for each vector boson mass individually.
From top to bottom, $m_b$=40, 80, and 120\,keV, respectively.
In each figure, three histograms are showing
events after the cumulative cuts 1-2 (dotted line),
cuts 1-3 (dashed line), and cut 1-4 (hatched histogram).
The effective livetime is 132.0 days 
and the target mass is 41\,kg.
The small number of events at the low energy region
in the final samples are due to lower efficiency of cut-4.
Efficiencies can be found in Tab.\ \ref{xmass:tab}.
For the simulated events, the dashed line (cuts 1-3) and
hatched histogram (cuts 1-4) are barely separated.
The coupling constants $\alpha'/\alpha$ assumed
in the simulation for 40, 80, and 120\,keV
are $2.0\times 10^{-24}$, $2.7\times 10^{-23}$, and $1.3\times 10^{-22}$,
respectively.
}
\label{xmass:data}
\end{figure}

\begin{figure}
\begin{center}
\includegraphics[width=11cm]{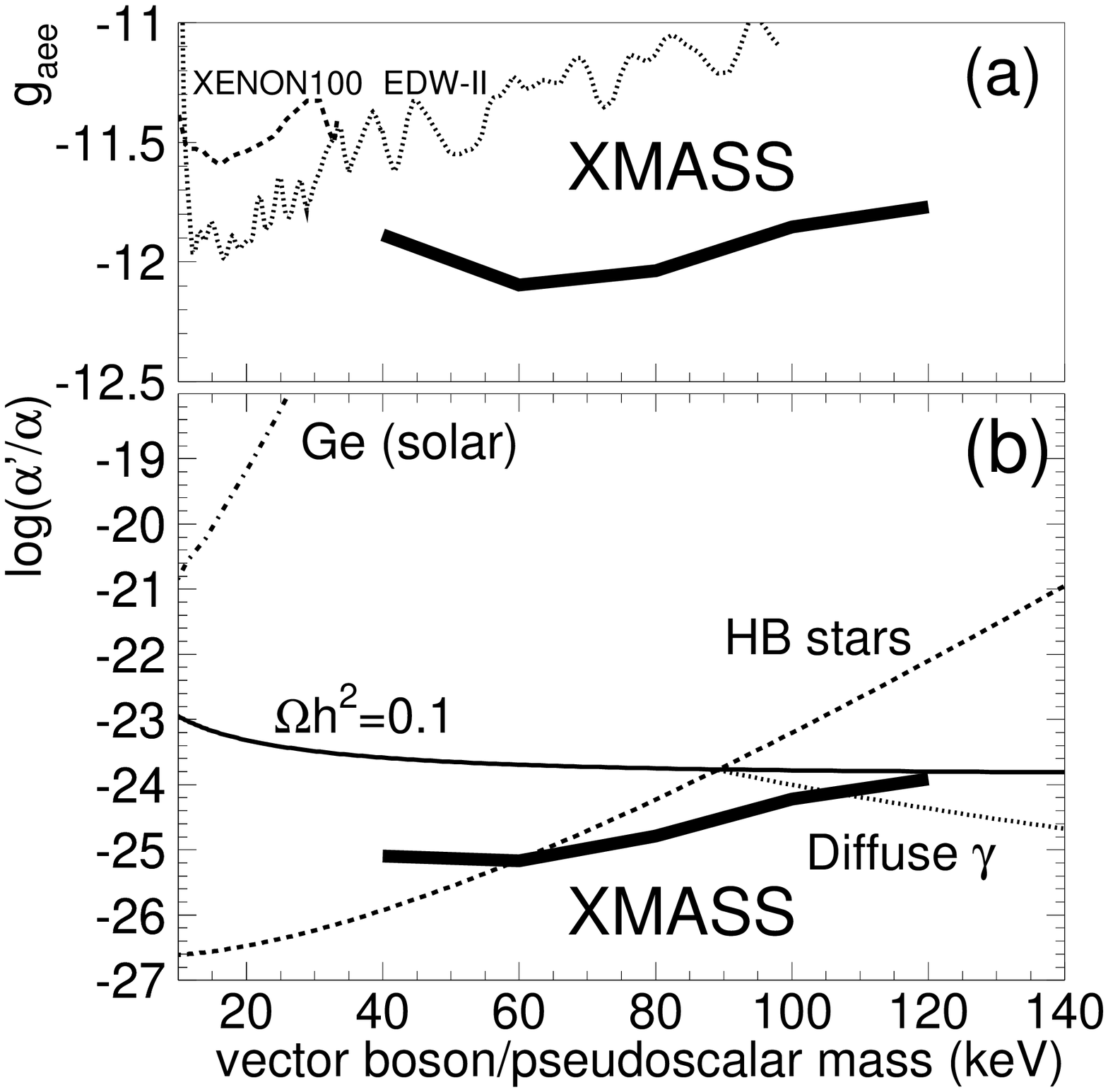}
\end{center}
\caption{Limits on coupling constants
for
(a) electrons and pseudoscalar bosons and
(b) electrons and vector bosons at 90\% CL
(thick solid line).
(a) Dashed line and dotted line correspond to
constraints obtained by EDELWEISS-II \cite{EDELWEISS} and XENON100 \cite{XENON100}. (b) 
The thin solid line corresponds to the coupling 
constant required to reproduce the observed
dark matter abundance including resonance effects
\protect{\cite{Pospelov2008,Redondo2009}}.
The dotted line and dashed line correspond to the upper limit
from the $\gamma$ ray background from 3$\gamma$ decays
in the Galaxy, and the constraint from the He-burning
lifetime in horizontal branch (HB) stars
\protect{\cite{Redondo2009}}.
The dash-dotted line shows an experimental 
constraint assuming production in the Sun \protect{\cite{Horvat}}.
}
\label{xmass:constraints}
\end{figure}

Figure \ref{xmass:data} shows data and simulated signal
after applying all the cuts, with the cuts optimized
as described in the previous section.
No significant
excess was seen in the data.
The remaining events stem mostly
from the radon daughter $^{214}$Pb. The amount
of radon was estimated by the observed rate of 
$^{214}$Bi-$^{214}$Po consecutive decays, 
and amounts to $8.2\pm 0.5$\,mBq \cite{XMASS2}.
Based on this rate we evaluated the expected
number of events in the signal window (see Tab.\ \ref{xmass:tab}).
This number is consistent with
the expectation except for the 40\,keV case,
where some leakage events caused by the radioactivity
on the inner surface may have not been rejected.
Since such background contributions are less certain,
we did not subtract such background when deriving upper limits.

Most of the
systematic error taken into account arise
from uncertainty in our cut efficiencies.
We have
used
$^{241}$Am data 
for $m_b=$ 40, 60, and 80\,keV and
$^{57}$Co data for $m_b=$ 100, and 120\,keV,
where the comparison between data and MC simulation is necessary.
For cut-1 systematic errors are negligible.
In cut-2 uncertainty for the reconstructed radius 
was estimated to be $\pm$1\,cm
by comparing the reconstructed vertex positions for 
data and simulation.
Changing the radius cut by $\pm$1\,cm, 
the resulting change in cut efficiency
ranges from
$\pm 13$\% to $\pm 17$\%,
depending on $m_b$.
For cut-3 the systematic uncertainties were evaluated 
from the difference in acceptance
between data and simulation,
and the systematic uncertainty
in modeling the scintillation
decay constants as a function of energy ($\pm$1.5\,ns).
The resulting systematic uncertainties for cut-3
range from $\pm 12$\% to $\pm 19$\%. 
For cut-4 we take the difference in acceptance between data
and simulation
($\pm$5\%). For the final event selection
using the $\pm$15\,keV$_{ee}$ signal window,
the majority of the systematic uncertainty comes
from the energy scale ($\pm 12$\%) and resolution ($\pm 5$\%).
Particularly, the scale uncertainty comes from 
the nonlinearity of the scintillation yield ($\pm$3.5\%),
position dependence ($\pm 2$\%),
and time variation ($\pm 3$\%).
All these systematic errors are used
in the evaluation of the detection efficiency uncertainty
listed
in Tab.\ \ref{xmass:tab}.

Dividing the number of events observed by the efficiency evaluated,
we derive a 90\% confidence level (CL) upper limit
on the number of bosons absorbed in the fiducial volume
without subtracting background evaluated.
In this calculation,
statistical and systematic uncertainties are added in quadrature.
Eqs.\ (2) and (4) were used to translate
this result into an upper limit on the
respective coupling constants, $\alpha'/\alpha$ or $g_{aee}$.
This result is given in Tab.\ \ref{xmass:tab},
and shown in Fig.~\ref{xmass:constraints}.
This is the first direct search for vector bosonic super-WIMPs
in this mass range.
In this range the present result excludes the possibility
for such WIMPs to constitute all of dark matter.
As can be seen in the figure,
the obtained limit also is
comparable to or
better than the current astrophysical constraints.
For pseudoscalar super-WIMPs coupling
the present limit improves significantly on previous results 
\cite{DAMA, CDMS, CoGeNT, EDELWEISS, XENON100}.
This significant improvement
was achieved exploiting the low background
in the detector at a level of $10^{-4}$\,kg$^{-1}$keV$_{ee}^{-1}$day$^{-1}$,
unprecedented in this energy range.

In summary
we searched in XMASS-I for 
signatures of bosonic super-WIMPs. 
In 165.9 days of data
with an effective livetime of 132.0 days
in a fiducial mass of 41\,kg,
no significant signal was observed
and stringent limits on the electron coupling
of bosonic super-WIMPs with masses in the 40-120\,keV range
were obtained.
For vector bosons the present experimental
limit excludes the possibility that vector super-WIMPs constitute all
the dark matter. The absence of the signal also provides the most
stringent direct constraint on the coupling constant of pseudoscalar 
dark matter to electrons.

\begin{acknowledgments}

We thank Naoki Yoshida for useful discussion on warm dark matter.
We gratefully acknowledge the cooperation of Kamioka Mining
and Smelting Company. 
This work was supported by the Japanese Ministry of Education,
Culture, Sports, Science and Technology, Grant-in-Aid
for Scientific Research, 
JSPS KAKENHI Grant Number, 19GS0204, and partially
by the National Research Foundation of Korea Grant funded
by the Korean Government (NRF-2011-220-C00006).
\end{acknowledgments}


\begin{thebibliography}{00}
\bibitem{Pospelov2008} M.~Pospelov, A.~Ritz, and M.~Voloshin, Phys. Rev. D {\bf 78,} 115012 (2008). 
\bibitem{Redondo2009} J.~Redondo, M.~Postma, J. Cosm. Astropart. Phys., {\bf 02,} 005 (2009).
\bibitem{WDM} K.~Markovi\v{c} and M.~Viel, Publ. Astron. Soc. of Aust., {\bf 31}, e006 (2014). 
\bibitem{PDG} J.~Beringer {\it et al.}, (Particle Data Group), Phys. Rev. D {\bf 86,} 010001 (2012).
\bibitem{XMASS2} K.~Abe {\it et al.} (XMASS collaboration), Nucl. Instr. Meth. A {\bf 716,} 78 (2013).
\bibitem{FN1} This photoelectron yield is smaller than 
the value reported in Ref.\ \cite{XMASS2, XMASS3, XMASS1}
since we changed a correction on the charge observed in the electronics.
This correction is within the uncertainty reported earlier, $\pm1.2$\,p.e./keV.
\bibitem{SKNIM} S.~Fukuda {\it et al.} (Super-Kamiokande Collaboration), Nucl. Instr. Meth. A {\bf 501,} 418 (2003).
\bibitem{XMASS3} K.~Abe {\it et al.} (XMASS collaboration), Phys. Lett B {\bf 724,} 46 (2013).
\bibitem{XMASS4} H.~Uchida {\it et al.} (XMASS collaboration), Phys. Theor. Exp. Phys., 
063C01 (2014).
\bibitem{LXeSpeed} S.~Nakamura {\it et al.}, in Proceedings of the Workshop on Ionization and Scintillation Counters and Their Uses, vol.~{\bf 27,} 2007.
\bibitem{XMASS1} K.~Abe {\it et al.} (XMASS collaboration), Phys. Lett B {\bf 719,} 78 (2013).
\bibitem{DAMA} R.~Bernabei {\it et al.} (DAMA collaboration), Int. J. Mod. Phys. A {\bf 21}, 1445 (2006).
\bibitem{CDMS} Z.~Ahmed {\it et al.} (CDMS collaboration), Phys. Rev. Lett. {\bf 103}, 141802 (2009).
\bibitem{CoGeNT} C.~E.~Aalseth {\it et al.} (CoGeNT collaboration), Phys. Rev. Lett. {\bf 101}, 251301 (2008).
\bibitem{EDELWEISS} E.~Armengaud {\it et al.} (EDELWEISS-II collaboration), JCAP {\bf 1311}, 067 (2013).
\bibitem{XENON100} E.~Aprile {\it et al.} (XENON100 collaboration), arXiv:1404.1455.
\bibitem{Horvat} R.~Horvat, D.~Kekez, M.~Kr\v{c}mar, Z.~Kre\v{c}ak, A.~Ljubi\v{c}i\'c, Phys. Lett. B {\bf 721,} 220 (2013).
\end{thebibliography}
\end{document}